\newcommand{\be}{\begin{equation}}\newcommand{\ee}{\end{equation}}
\newcommand{\bea}{\begin{eqnarray}}\newcommand{\eea}{\end{eqnarray}}
\newcommand{\nn}{\nonumber}\newcommand{\p}[1]{(\ref{#1})}
\newcommand{\lb}[1]{\label{#1}}
\newcommand\pdap{\partial^{\dot{\alpha}+}}

\newcommand\pdam{\partial^{\dot{\alpha}-}}
\newcommand\pdbm{\partial^{\dot{\beta}-}}

\newcommand\pab{\partial_{\alpha\dot{\beta}}}

\newcommand\adb{{\alpha\dot{\beta}}}
\newcommand\bdg{{\beta\dot{\gamma}}}
\newcommand\gdr{{\gamma\dot{\rho}}}

\newcommand\da{{\dot{\alpha}}}
\newcommand\db{{\dot{\beta}}}
\newcommand\dg{{\dot{\gamma}}}
\newcommand\dr{{\dot{\rho}}}
\newcommand\ds{{\dot{\sigma}}}

\newcommand\dap{{\dot{\alpha}+}}
\newcommand\dbp{{\dot{\beta}+}}

\newcommand\dam{{\dot{\alpha}-}}

\newcommand\T{\mbox{Tr}}
\newcommand\s{\scriptscriptstyle}

\newcommand\C{{\s C}}

\newcommand{\pp}{{\s ++}}
\newcommand{\m}{{\s --}}
\newcommand{\0}{{\s 0}}
\newcommand{\2}{{\s 2}}

\newcommand{\4}{{\s 4}}

\newcommand{\Dp}{D^{\pp}}
\newcommand{\Dm}{D^{\m}}
\newcommand{\Vp}{V^{\pp}}
\newcommand{\Vm}{V^{\m}}

\newcommand{\dpp}{\partial^\pp}
\newcommand{\dm}{\partial^\m}

\documentstyle[12pt]{article}

\begin{document}
\renewcommand{\thefootnote}{\fnsymbol{footnote}}
\begin{titlepage}
\begin{flushright}
{December 1998 }\\
{ITP-UH-36/98}\\
{JINR E2-98-382}\\
{hep-th/9812244}
\end{flushright}
\vspace{2cm}
\begin{center}
{\Large\bf MANIFESTLY INVARIANT ACTIONS } \\[10pt]
{\Large\bf FOR HARMONIC SELF-DUAL GAUGE THEORY} \\
\vspace{2cm}
{\large\bf Olaf Lechtenfeld}\\
\vspace{0.2cm}
{\it Institut f\"ur Theoretische Physik, Universit\"at Hannover,
30167 Hannover, Germany} \\
{\it lechtenf@itp.uni-hannover.de}\\
\vspace{1cm}
{\bf and} \\
\vspace{1cm}
{\large\bf \mbox{\phantom{xx}} Boris Zupnik~\footnote{
On leave of absence from: \\ \phantom{XX}
{\it Institute of Applied Physics, Tashkent State University,
Tashkent 700095, Uzbekistan}}
}\\
\vspace{0.2cm}
{\it Bogoliubov Laboratory of Theoretical Physics, Joint Institute for
Nuclear Research, 141980 Dubna, Moscow Region, Russia\\
zupnik@thsun1.jinr.ru}\\
\end{center}
\vspace{2cm}

\begin{abstract}
We discuss alternative descriptions of four-dimensional self-dual Yang-Mills
fields in harmonic space with additional commuting spinor coordinates.
In particular, the linear analyticity equation and nonlinear covariant
harmonic-field equations are studied. A covariant harmonic field can be treated
as an infinite set of ordinary four-dimensional fields with higher spins.
We analyze different constructions of invariant harmonic-field actions
corresponding to the self-dual harmonic equations.
\end{abstract}

\end{titlepage}

\renewcommand{\thefootnote}{\arabic{footnote}}
\setcounter{footnote}0
\setcounter{equation}0
\section{Introduction}
\hspace{0.5cm}
It is well known that the self-dual Yang-Mills (SDYM) equations in
four-dimensional Euclidean space $E_4$ have equivalent
sigma-model-type representations \cite{Ya,Le}. In contrast with
the original self-duality equation, these representations are not
manifestly invariant under $SO(4)\simeq SU(2)_L\times SU(2)_R$ rotations.
These remarkable $4D$ field models have been intensively
studied \cite{Do,NS,LMNS,Ke}, and their applications as effective field models
for the open $N{=}2$ string have also been discussed \cite{OV,Ma,Si}.
Different sigma-model representations of the SDYM equations are equivalent
classically; however, the consistency and compatibility of these models
in the quantum region are not evident.

The problem of constructing an $SO(4)$-invariant action for
self-dual gauge fields was also investigated.
The simplest invariant action for the SDYM equations contains
propagating Lagrange multiplier \cite{CS}. An improved
invariant action based on an infinite set of non-propagating
auxiliary fields has been considered in ref.\cite{BH}.

The geometric description of the self-dual solutions is connected with
twistor methods \cite{BZ,Wa}. We shall use the harmonic approach to the 
SDYM equations \cite{GIKOS,GIOS1,GIOS23}, which is a covariant version of
the twistor method. The (anti)self-dual solutions are described
in harmonic-twistor space using additional harmonic coordinates
$u^\pm_\alpha$ which are $SU(2)_L$ spinors.
The space of  self-dual fields is parametrized by the set of
harmonic connections $\Vp(x,u)$ satisfying a linear analyticity equation.
Fields in harmonic space are equivalent to infinite sets of ordinary
$4D$ fields (harmonic components) of higher spins (helicities).
The analyticity condition on $\Vp$ translates to an
infinite set of  linear equations for the covariant component fields.
Note that analogous infinite multiplets of fields appear
in the bosonic variant of superspace proposed in ref.\cite{DL} for
describing the effective field theory of the $N{=}2$ string.

After briefly reviewing non-covariant formulations of self-dual Yang-Mills
theory in sect.\ref{B}, we shall consider in sect.\ref{C}
harmonic representations of the SDYM equations.
These generalize the non-covariant self-dual equations \cite{Ya,Le}.
We then discuss real forms of the harmonic fields in different representations.
In sect.\ref{F}, the dual harmonic connection $\Vm(x,u)$ is used
to propose a harmonic generalization of the Leznov equation.
Also, a harmonic analog of the Yang equation can be
written in terms of a harmonic field $L(x,u)$ which contains odd spins only.
On-shell all these equations again produce the SDYM equations for the
Yang-Mills field, which can depend on the harmonic coordinates.
However, one can entertain the speculation that
the higher-spin harmonic components propagate off-shell and generate
new effects in the quantum theory.

An off-shell description of self-dual harmonic fields has been proposed
in ref.\cite{KS} but was shown to lead to a perturbatively trivial
$S$-matrix \cite{MOY}.
In  sect.\ref{D}, we discuss alternative constructions of invariant
harmonic actions using different harmonic-field variables.
One can consider a modification of the simplest action for the analyticity
equation by adding polynomial terms which produce on-shell relations
between the Lagrange multiplier and the prepotential. It is possible that such
a modification allows for nontrivial quantum corrections, although
the precise consequences for quantization are unclear.

Second-order self-dual harmonic equations can be treated as nonlinear
generalizations of the ordinary Laplace equation and can be quantized
with the help of  standard methods. In sect.\ref{E}, we consider the
quadratic part of an invariant action for the harmonized Leznov equation,
since we do not know the exact form of the corresponding interactions.
The most simple action is constructed for the harmonic analog of the
Yang equation.

Useful relations for $SU(2)_L/U(1)$ harmonics are collected in an appendix.

Possible applications and quantum calculations with harmonic self-dual actions
are not considered in this paper. However, we hope that the harmonic approach
will be useful for revealing the role of self-dual fields in quantum field
theory.

\setcounter{equation}{0}
\section{\label{B} Non-covariant formulations of self-dual theories}
\hspace{0.5cm}
The four-dimensional equations for self-dual Yang-Mills (SDYM) fields are
consistent in $4D$ Euclidean space $E_4$
as well as in $4D$ Kleinian space $K_4$ (of signature (2,2)).
The spinor covering groups
for these spaces are $SU(2)_L\times SU(2)_R$ and $SL(2,R)_L\times SL(2,R)_R$,
respectively. Since we shall consider only the formal problem of constructing
actions for SDYM theory, we may restrict our analysis to the Euclidean
case. Let $\alpha,\beta\ldots$ and $\da,\db \ldots$ be 2-spinor
indices of the $SU(2)_L$ and $SU(2)_R$ groups, respectively, and
denote the Euclidean Weyl matrices by
\be
(\sigma^m)_\adb \qquad {\rm and} \qquad
(\bar{\sigma}^m)^{\db\alpha}=\varepsilon^{\alpha\rho}\varepsilon^{\db\dg}
(\sigma^m)_{\rho\dg}\lb{B1} \quad.
\ee
Bispinor representations for the 4D coordinates and partial derivatives are
\be
x^\adb={1\over\sqrt{2}}(\bar{\sigma}_m)^{\db\alpha}x^m \qquad{\rm and}\qquad
\pab ={1\over\sqrt{2}}(\sigma^m)_\adb\partial_m \quad. \lb{B2}
\ee

The (anti)self-duality equation for the non-Abelian gauge field $A_\adb=
(\sigma^m)_\adb A_m $ has the following form:
\be
[\nabla_\adb, \nabla_\gdr]=\varepsilon_{\alpha\gamma}F_{\db\dr} \quad,
\lb{B3}
\ee
where $\nabla_\adb=\pab +A_\adb $ is the covariant derivative.
It is useful to rewrite these equations in an $SU(2)_L$ non-covariant form,
\bea
&&[\nabla_{{\s1}\da},\nabla_{{\s1}\db}] = 0 \quad,\lb{B4}\\
&&[\nabla_{{\s2}\da},\nabla_{{\s2}\db}] = 0 \quad,\lb{B5} \\
&&\varepsilon^{\da\db}[\nabla_{{\s1}\da},\nabla_{{\s2}\db}] = 0 \quad.\lb{B6}
\eea

The Yang ansatz \cite{Ya} solves the first two equations of this system via
\be
A_{{\s1}\db}=h^{-1}\partial_{{\s1}\db}h \qquad{\rm and}\qquad
A_{{\s2}\db}=\bar{h}\partial_{{\s2}\db}\bar{h}^{-1} \quad, \lb{B7}
\ee
where $h$ and $\bar{h}$ are independent $SL(N,C)$ matrices.
The reality condition on the gauge potential reads
\be
(A_{{\s1}\db})^\dagger =-A_{\s2}^\db \quad\mbox{in}\;E_4 \qquad\mbox{or}\qquad
(A_{\adb})^\dagger =-A_{\adb} \quad\mbox{in}\;K_4 \quad.
\lb{B7b}
\ee
Consequently, one must choose $\bar{h}=h^\dagger$ in the $E_4$ case but
take independent unitary matrices $h$ and $\bar{h}$ in $K_4$.
The remaining equation \p{B6} within this ansatz,
\be
\partial_{\s1}^\da \Bigl[(h\bar{h})\partial_{{\s2}\da}(h\bar{h})^{-1}\Bigr]=0
\quad,\lb{B8}
\ee
resembles the equation for the 2D principal chiral model.
In terms of the matrix variable $l=\mbox{ln}(h\bar{h})$,
this equation can be obtained from the action \cite{Do,NS}
\be
S[l]=\int\!\! d^{\4}\!x\!\int\limits^{1}_{0}\!\! ds\;\T\;l\;\partial_{\s1}^\da
[e^{sl}\;\partial_{{\s2}\da}e^{-sl}]
=\int\!\! d^{\4}\!x\;\T\Bigl( l\; \partial_{\s1}^\da \partial_{{\s2}\da}\;l
-{\textstyle{1\over3}}l\;[~\partial_{\s1}^\da l~,~ \partial_{{\s2}\da}l~]
+O(l^4) \Bigr) \quad,\lb{B9}
\ee
where $s$ is some auxiliary parameter. Note that this nonlinear
action for SDYM theory is not invariant with respect to $SU(2)_L$.

The alternative Leznov ansatz \cite{Le} for the SDYM equations reads
\be
A_{{\s1}\da}=0 \qquad{\rm and}\qquad A_{{\s2}\da}=\partial_{{\s1}\da} A
\quad, \lb{B10}
\ee
where $A$ is some Lie-algebra-valued function.
This ansatz trivially solves eq.\p{B6},
but now eq.\p{B5} becomes dynamical,
\be
\partial_{\s1}^\da \Bigl(\partial_{{\s2}\da} A + {\textstyle{1\over2}}
[A~,~\partial_{{\s1}\da} A]\Bigr)=0 \quad.\lb{B11}
\ee
Clearly, it is equivalent to the following first-order equation \cite{Le}
\be
\partial_{{\s2}\da} A + {1\over2}[~A~,~\partial_{{\s1}\da} A~]=
\partial_{{\s1}\da}\Omega \quad,\lb{B11b}
\ee
where $\Omega$ is some arbitrary function.

It is interesting that Yang's equation \p{B8} is non-polynomial in $l$ while
Leznov's equation \p{B11} is quadratic in $A$.
The simplest corresponding action is cubic,
\be
S[A]=\int\!\! d^{\4}\!x\;\T \Bigl( A\partial_{\s1}^\da \partial_{{\s2}\da} A
+ {\textstyle{1\over6}} A\;
[~\partial_{\s1}^\da A~,~\partial_{{\s1}\da} A~]\Bigr)=0 \quad.\lb{B12}
\ee

Alternatively, the first-order equation \p{B11b} can be obtained from
\be
S(P,A,\Omega)=\int\!\! d^{\4}\!x\; \T\; P^\da \Bigl\{ \partial_{{\s2}\da} A +
{\textstyle{{1\over2}}}[~A~,~\partial_{{\s1}\da} A~] -
\partial_{{\s1}\da}\Omega \Bigr\} + S^\prime(A) \quad,\lb{B14}
\ee
where $P^\da$ is a dynamical Lagrange multiplier.
The additional term $S^\prime(A)$ only affects the equation
for $P^\da$ and does not modify eq.\p{B11b}.

The two given representations of the on-shell self-dual gauge field
are related by
\be
e^l \partial_{{\s2}\da} e^{-l}=\partial_{{\s1}\da} A \quad. \lb{B13}
\ee
This relation will not be valid off-shell, and so the quantum relation between
different representations requires further analysis.

\setcounter{equation}{0}
\section{\label{C} Twistor-harmonic solutions of SDYM equation}
\hspace{0.5cm}
The connection of the SDYM equation with the inverse-scattering method has
been  considered in ref.\cite{BZ}. The corresponding equation of the
auxiliary linear problem has the following form:
\be
(\nabla_{{\s1}\da} +\lambda\nabla_{{\s2}\da})\ g(\lambda,x)=0 \quad,\lb{invsc}
\ee
where $\lambda$ is the complex spectral parameter, and $g(\lambda,x)$ is
the auxiliary matrix function, usually chosen as a meromorphic
function of $\lambda$ with the specific rule of conjugation
\be
g^\dagger(-1/\bar{\lambda})=g^{-1}(\lambda) \quad.\lb{conj}
\ee

In the twistor interpretation of the SDYM equation \cite{Wa} the projective
line $CP(1)$ is described by complex coordinates $\eta^\alpha$. The
anti-self-dual solutions can be written with the help of a flat connection
in this special direction,
\be
\eta^\alpha \nabla_\adb=g^{-1}(\eta,x)\eta^\alpha\pab g(\eta,x)\quad.\lb{C0}
\ee
The harmonic description of the SDYM equation \cite{GIOS23}
can be viewed as a covariant version of the twistor approach,
using the isomorphism between $CP(1)$ and the coset $SU(2)_L/U(1)$.
The harmonics
$u^\pm_\alpha$ can be introduced as additional coordinates of the
harmonic-twistor space. They transform covariantly with respect to
$SU(2)_L$ and an extra $U(1)$ \cite{GIKOS,GIOS1}. In the special
complex parametrization of harmonics \p{G18} the $SU(2)_L$ transformations
correspond to rational transformations of the complex plane.
It should be stressed that the practicioners of the harmonic approach treat
harmonics as global coordinates on the sphere $SU(2)_L/U(1)$ and
employ special representations only when comparing with other methods.
All covariant functions of harmonics possess regular series expansions
on the sphere, and conservation of the extra $U(1)$ charge is a necessary
condition in this approach.

Using the
$SU(2)_L$-invariant harmonic projections of the covariant derivatives,
\be
(\nabla_\C)_\db^+=u^{+\alpha}\nabla_\adb=\partial_\db^+ + C_\db^+ \quad,\qquad
(\nabla_\C)_\db^-=u^{-\alpha}\nabla_\adb=\partial_\db^- + C_\db^- \quad,
\lb{C1}
\ee
one can obtain the harmonic decomposition of the SDYM equation \p{B3},
\bea
&& F^\pp_{\da\db}\equiv [(\nabla_\C)_\da^+,(\nabla_\C)_\db^+]=0\quad,\lb{C2}\\
&& F^\m_{\da\db}\equiv [(\nabla_\C)_\da^-,(\nabla_\C)_\db^-]=0\quad,\lb{C3}\\
&& F^{\s+-}\equiv [(\nabla_\C)^\dap,(\nabla_\C)_\da^-]=0\quad.\lb{C4}
\eea

In addition to these equations of dimension $d{=}{-}2$, we may use the
following $d{=}{-}1$ equations of the harmonic approach:
\bea
&&  F^{+++}_\da\equiv[(\nabla_\C)^\pp , (\nabla_\C)_\da^+]=0 \quad,\qquad
F^+_\da\equiv[(\nabla_\C)^\pp,(\nabla_\C)_\da^-]-(\nabla_\C)_\da^+=0 \quad,
\lb{C5}\\
&&  F^{---}_\da\equiv[(\nabla_\C)^\m, (\nabla_\C)_\da^-]=0 \quad,\qquad
F^-_\da\equiv[(\nabla_\C)^\m,(\nabla_\C)_\da^+]-(\nabla_\C)_\da^-=0 \quad.
\lb{C6}
\eea
where covariant harmonic derivatives $\nabla^{\s\pm\pm}$ are introduced.
Until now, we worked in the so-called central basis (CB) with $u$-independent
gauge group and flat harmonic derivatives
$(\nabla_\C)^{\s\pm\pm}=\partial^{\s\pm\pm}$.
The properties of harmonics and partial harmonic
derivatives \cite{GIKOS,GIOS1} are considered in the appendix.

One can define two types of conjugations for the harmonic matrix functions,
namely $C(u)\rightarrow \bar{C}(u)$ and $C(u)\rightarrow \widetilde{C}(u)$
according to (\ref{G5}) and (\ref{G6}).
In the case of a unitary gauge group, the harmonic projections ${C_\db^\pm}$
of the anti-Hermitean CB connections $A_{\adb}$ satisfy the following reality
conditions with respect to these conjugations,
\be
\overline{C_\db^\pm}=\pm C^{\db\mp} \quad,\qquad
\widetilde{C}_\db^\pm=-C^{\db\pm} \quad.\lb{C6b}
\ee

The general harmonic solution of equation \p{C2} has the following form:
\be
C_\db^+=g^{-1}\partial_\db^+ g \quad,\lb{C7}
\ee
where the bridge matrix $g(x,u)$ satisfies the analyticity condition
\be
[\partial_\db^+,\Vp]=0 \qquad{\rm with}\qquad \Vp=-(\dpp g)g^{-1}\quad.\lb{C8}
\ee

The bridge $g$ provides a transformation between the geometric
quantities of the CB and a new analytic basis (AB), defined by the relations
\bea
&& \nabla = g (\nabla_\C) g^{-1} \quad,\qquad
{\cal F}=g F_\C g^{-1}\quad,\lb{C8b}\\
&&\nabla_\db^+ =\partial_\db^+ \quad,\qquad
\nabla^\pp=\dpp + \Vp \quad,\lb{C9}\\
&&\nabla_\db^- =\partial_\db^- + A_\db^- \quad,\qquad
A_\db^-=-\partial_\db^+\Vm \quad,\lb{C10}\\
&&\nabla^\m=\Dm +\Vm \quad,\qquad
\dpp\Vm -\dm\Vp+[\Vp,\Vm]=0 \quad.\lb{C11}
\eea
The gauge group in the AB is analytic,
\be
\delta V^{\s\pm\pm}=\Dm \omega +[V^{\s\pm\pm} , \omega] \quad,\qquad
\partial^+_\da \omega=0 \quad,\lb{Cab}
\ee
while the bridge matrix enjoys a mixed transformation law
\be
\delta g= \omega g - g\tau \quad,\lb{Cbr}
\ee
where $\tau$ is the CB gauge parameter.

The standard harmonic SDYM solution \cite{GIOS23} uses the analytic
connection $\Vp$ as initial data and reduces the full problem to solving
the linear harmonic equation
\be
(\dpp + \Vp)\ g =0 \quad.\lb{C12}
\ee
In the representation \p{G19} one has
$\dpp g \sim \partial g/\partial \bar{\lambda}$,
so this equation generalizes the meromorphicity condition on the auxiliary
matrix function in the standard twistor approach. (See, however,
ref.\cite{CN} for an analogous equation in the twistor formalism.)
The harmonic equation possesses
a simple one-instanton solution \cite{GIOS23} regular in harmonics.
Note that the reality condition becomes simply
\be
\widetilde{g}(u,x)=g^{-1}(u,x)\quad,\qquad\widetilde{V}^\pp=-\Vp\quad.\lb{real}
\ee
The other conjugation is not used in this analytic representation of
the self-dual solutions.

It is useful to introduce analytic coordinates and derivatives
\be
x^{\db\pm}=u^\pm_\alpha x^\adb \quad,\qquad
\partial_\db^\pm=u^{\pm\alpha}\pab \quad,\qquad
\partial_\db^\pm x^{\da\mp}=\pm\delta^\da_\db \quad. \lb{C12b}
\ee
Analytic functions do not depend on the $x^\dam$ coordinates,
like $\Vp=\Vp(x^+,u^\pm)$. Note that in analytic coordinates we should
use the covariant harmonic derivatives
\be
D^{\s\pm\pm}=\partial^{\s\pm\pm} \pm x^{\da\pm}\partial^\pm_\da \quad,\qquad
[\Dp,\pdam]=\pdap \quad,\qquad [\Dm,\pdap]=\pdam \quad.\lb{C12c}
\ee

The anti-self-dual gauge field in the central basis and corresponding
field strength can be written in terms of the harmonic quantities as
\bea
&&A_{\adb}(\Vp)=u^-_\alpha C^+_\db -u^+_\alpha C^-_\db=
g^{-1}(\pab - u^+_\alpha A_\db^-)\ g \quad,\lb{C12d}\\
&& F_{\da\db}(\Vp)=g^{-1}\partial^+_\da\partial^+_\db \Vm g\quad.\lb{C12e}
\eea
This wide class of self-dual fields does not depend manifestly on harmonics,
\be
\dpp A_{\adb}(\Vp)=0 \quad,\qquad \dpp F_{\da\db}(\Vp)=0 \quad, \lb{C12f}
\ee
if eqs.(\ref{C8})-(\ref{C12}) are satisfied. In the following section we shall
discuss SDYM configurations $A_{\adb}$ which depend parametrically on the
harmonics and cannot be expressed via the analytic connection.

The definition of the analytic basis is formally independent of the analyticity
condition \p{C8}, so we may consider the nonlinear zero-curvature equation
\be
[\nabla_\db^-,\nabla^\m]\equiv {\cal F}_\db^{---}(\Vm)=\partial_\db^-\Vm +
\Dm\partial_\db^+\Vm +[\Vm,\partial_\db^+\Vm]=0 \lb{C14}
\ee
as an alternative analyticity condition in the AB.
Equation \p{C14} is equivalent to the linear equation \p{C8}, and its
perturbative solution can be written in terms of the analytic prepotential
$\Vp$ \cite{Z1} as
\be
\Vm(x,u)=\sum\limits^{\infty}_{n=1} (-1)^n \int\!\! du_{\s1} \ldots du_n
\frac{\Vp(x,u_1 )\ldots \Vp(x,u_n )}
{(u^+ u^+_{\s1})(u^+_{\s1} u^+_{\s2})\ldots (u^+_n u^+ )}  \label{C14b}
\ee
where  $(u^+_{\s1} u^+_{\s2} )^{-1}$ is the harmonic distribution
\cite{GIOS1}.
Conversely, one may treat $\Vm$ as a basic harmonic field
and obtain the connection $\Vp$ as the solution of \p{C11}.
The bridge matrix of this formulation satisfies the equation
\be
(\dm + \Vm)g =0 \quad.\lb{C13}
\ee

\setcounter{equation}{0}
\section{\lb{F} Generalized harmonic self-dual equations}
\hspace{0.5cm}
The partial derivative of the alternative analyticity equation \p{C14}
produces a harmonic generalization of the Leznov equation \p{B11},
\be
[\partial^\dap, {\cal F}_\da^{---}(\Vm)]=
2\partial^\dap \partial_\da^-\Vm
+ [\partial^{\da+}\Vm ,\partial_\da^+\Vm]=0 \quad.\lb{F1}
\ee
Note that this second-order equation is invariant under the additional
transformation $\delta \Vm=\omega^\m$ with an analytic parameter $\omega^\m$.
In distinction with \p{C14}, this equation does not contain harmonic
derivatives, and so the dependence of $\Vp$ on harmonics is merely parametric.
In the non-covariant approach, one can treat the Leznov field $A$ \p{B11}
as a special configuration of the covariant harmonic field $\Vm$.

Note that the second-order equation \p{F1} is equivalent to the
first-order equation
\be
2\partial_\da^-\Vm + [\Vm ,\partial_\da^+\Vm]=\partial_\da^+\Omega^{----}
\quad,\lb{F2}
\ee
where $\Omega^{----}(x,u)$ is an arbitrary function.
Thus, the second-order harmonic equation \p{F1} is in some sense more general
than the first-order equations \p{C8} or \p{C14}. The general solutions $\Vm$
of \p{F1} correspond to non-analytic connections $\Vp$. Using the bridge
matrix $g$ \p{C13} for these solutions one can obtain the CB-connection
$A_{\adb}(x,u)$ \p{C12d} which manifestly depends on harmonics in this case.
Nevertheless, the study of this representation  can be useful for the
construction of self-dual gauge fields.

It should be underlined that the solutions of the $4D$ Laplace equation,
arising as the linear approximation of the self-dual equations
(\ref{B8}) and (\ref{B11}), have the general harmonic-twistor
representation
\be
f=\int\!\! du\; F(x^\dap,~u^\pm_\alpha) \quad,\qquad \Box f=0 \quad,\lb{F4}
\ee
where $F$ is an analytic harmonic function. This representation connects
solutions of the non-covariant self-dual equations and their harmonic
generalizations with the space of analytic functions.
Each solution of the abelian equation
$\pdap A^+_\da=0$
can be written in the different equivalent forms
\be
A^+_\da=\partial_\da^+ v_\0(x,u)=\partial_\da^-a^\pp_\0(x,u) \quad,\lb{eqv}
\ee
with $a^\pp_\0(x,u)$ solving the $4D$-Laplace equation.
The proof follows from eq.\p{F4} 
or can be obtained by the expansion in $x^\dam$.
Let us consider the harmonic generalization of the Yang representation in
the central basis
\be
C_\db^+=g^{-1}\partial_\db^+ g \quad,\qquad
C_\db^-=(-\partial_\db^-\bar{g})\bar{g}^{-1} \quad,  \lb{F5}
\ee
where the matrix $\bar{g}$ is defined via transposition and
ordinary conjugation \p{G5}. The simultaneous use of two different
conjugations is an interesting peculiarity of this harmonic representation.
Using the transformation $A^\pm_\da=g(\partial^\pm_\da +C^\pm_\da) g^{-1}$
to the analytic basis, one can obtain the following representation:
\be
A_\db^+=0 \quad,\qquad
A_\db^-= -(\partial_\db^- g\bar{g})(g\bar{g})^{-1} \quad.\lb{F6}
\ee
Now introduce the new matrix variable $L(x,u)$
\be
g\bar{g}=e^L\quad,\qquad L=v+\bar{v}+{1\over2}[v,\bar{v}]+\ldots\quad,\lb{F7}
\ee
where $v=\ln g$.
This matrix satisfies specific reality properties
with respect to the two conjugations,
\be
\widetilde{L}(x,u)=-L(x,u) \quad,\qquad \bar{L}(x,u)=L(x,u)\quad.\lb{F8}
\ee
Note that these conditions are compatible for the following harmonic
decomposition
\be
L(x,u)=\sum\limits^{\infty}_{k=0}(u^{+(2k+1)}
u^{-(2k+1)})_{\alpha_1\cdots\alpha_{4k+2}} L^{\alpha_1\cdots\alpha_{4k+2}}(x)
\quad.\lb{F9}
\ee
It is not difficult to show that these $U(1)$-neutral harmonics with odd 
spins $2k+1$ are imaginary or real with respect to the different 
conjugations, and so all
fields $L^{\alpha_1\cdots\alpha_{4k+2}}(x)$ are anti-Hermitean.
This representation trivially solves eqs.(\ref{C2}) and (\ref{C3})
and produces the harmonic generalization of the Yang equation \p{B8}:
\be
\partial^\dap \Bigl[e^L(\partial_\da^- e^{-L})\Bigr]=0 \quad.\lb{F10}
\ee

Finally, let us relate the harmonic
SDYM configurations in different representations. Equation \p{F7}
connects the field $L$ with the solutions of the harmonic bridge equation
\p{C13} corresponding to the self-dual field $\Vm$.
The inverse construction of the fields $g$ and $\Vm$ in terms of $L$ is not
so straightforward. One considers the equation
\be
 e^L[\partial_\db^- e^{-L}]= -\partial_\db^+\Vm \quad,
\lb{F11}
\ee
which is the non-abelian analog of the relation \p{eqv}.
The explicit relation between $\Vm$ and $L$ has the following nonlocal form:
\be
\Vm(L)={1\over\Box}\partial^{\db-}[e^L(\partial_\db^- e^{-L})]+\Lambda^\m
\quad,\lb{F12}
\ee
where $\Box=\partial^\dbp\partial_\db^-$ and $\Lambda^\m$ is an analytic
function. This expression can be used in the harmonic equation \p{C13}
for the construction of the bridge
$g(L)$ corresponding to the solution $L$ of eq.\p{F10}.

Thus, the harmonic space may be employed to formulate various
covariant equations which are equivalent to the self-duality equation
in ordinary space. These formulations allow us to describe more
precisely the structure of the moduli space for the SDYM equation.

\setcounter{equation}{0}
\section{\lb{D}  Invariant actions for the analyticity equations }
\hspace{0.5cm}
In the non-covariant formalisms of
SDYM theory one may analyze the actions for the fields $l$ \p{B9} or
$A$ \p{B12} off-shell and construct the corresponding functional integral.
These field theories are used for the analysis of $N{=}2$ string amplitudes
and for calculations of the Yang-Mills $S$-matrix in the self-dual sector.

Attempts to construct an invariant action for the self-duality
equation lead to the problem of doubling of states or even to the appearance
of an infinite number of auxiliary fields \cite{BH}. An infinite number of
physical states also arise in the effective field theory
for the $N{=}2$ string \cite{DL}.

In the harmonic approach the analytic prepotential $\Vp$ or its
alternative cousins $\Vm$ and $L$ are equivalent
to the on-shell self-dual fields in Euclidean space.
We may try to treat the covariant harmonic equations as basic field
equations in the extended harmonic space $E_4\times S_2$ and use
unconstrained  harmonic fields to define the theory off-shell.
In contrast with the Kaluza-Klein approach,
the harmonic-field theory identifies the symmetry group of $S^2$ with
the $SU(2)_L$ subgroup of $SO(4)$.
It is evident that this covariant harmonic theory contains an infinite
number of ordinary $4D$ fields by construction, but its connection with
the non-covariant theories containing a finite number of fields is not clear.
Here we shall analyze invariant actions
for  harmonic equations, hoping to find
a consistent  description of the nontrivial interactions of
higher-spin fields.

In order to preserve manifest covariance off-shell we treat one
of the dimensionless harmonic fields as the basic field variable in the action.
The purely harmonic equations (\ref{C11}) and (\ref{C12}) with $d{=}0$
are independent of the analyticity conditions, and so we can use
off-shell the relations
\be
V^{\s\pm\pm}=-(\partial^{\s\pm\pm} e^v)e^{-v} \quad.\lb{D0}
\ee
For a unitary gauge group one should also keep off-shell the conjugation rules
\be
\widetilde{V}^{\s\pm\pm}=-V^{\s\pm\pm}\quad,\qquad\widetilde{v}=-v\quad.\lb{D1}
\ee
We covariantly decompose the off-shell fields
into an infinite set of fields with arbitrary spins $k$,
\bea
&&V^{\s\pm\pm}(x,u)=\sum\limits^{\infty}_{k=1}
(u^{\pm(k+1)} u^{\mp(k-1)})_{\alpha_1\cdots\alpha_{2k}}
V^{\alpha_1\cdots\alpha_{2k}}_k(x) \quad, \lb{D2}\\
&& v(x,u)=v_\0(x)+\sum\limits^{\infty}_{k=1}
(u^{+k} u^{-k})_{\alpha_1\cdots\alpha_{2k}}
v^{\alpha_1\cdots\alpha_{2k}}_k(x) \quad.\lb{D3}
\eea

Note that these harmonic component fields radically differ from the infinite
set of bispinor auxiliary fields $G^{\alpha\beta}_n$ of ref.\cite{BH}, which
are covariant with respect to the ordinary gauge group and have $d{=}{-}2$.
On the other hand, the invariant self-dual action of ref.\cite{CS} contains
only one additional propagating field $G^{\alpha\beta}_\0$,
\be
\int\!\! d^{\4}\!x\;\T\;G^{\alpha\beta}_\0
\Bigl(\partial_\alpha^\dg A_\bdg + [A_\alpha^\dg,A_\bdg]\Bigr) \quad.\lb{D4}
\ee

A general formulation of the harmonic gauge theory in the AB requires
a choice of basic parametrization of the connection $A^-_\da$
in terms of the independent harmonic variable $\Vp,~\Vm$, or $v$.
The analytic on-shell gauge group
of the harmonic self-dual equations is unessential for the off-shell
harmonic fields, and we do not know the appropriate generalization of the
gauge group in the harmonic space. However, we shall present
several possible actions for the harmonic equations considered in sect.\ref{C}.
The off-shell geometry in the analytic basis automatically
guarantees the relation ${\cal F}^\pp_{\da\db}=0$, but
we might want to include additional off-shell constraints.
The formal use of the relations \p{C12d} determines
the gauge field, which depends on the harmonics off-shell.

We should like to discuss shortly the problem of truncating the
infinite set of harmonic component fields satisfying the various types of
self-duality equations. The linear analyticity equation
$\partial^+_\da \Vp=0$ is compatible with an arbitrary truncation of the
harmonic expansion  of $\Vp$.  Nevertheless, it is difficult to use
truncated harmonic fields in the action. The nonlinear equations
\p{F1} and \p{F10} are not consistent with simple truncations
of the harmonic expansion for $\Vm$ and $L$.

The bilinear action with the Lagrange multiplier $P^{\da---}(x,u)$
\be
S_P=\int\!\! dx\,du\;\T\;P^{\da---}\partial^+_\da \Vp \lb{D5}
\ee
is trivial if one treats $\Vp(x,u)$ as the basic off-shell field.
The authors of ref.\cite{KS} have instead used the field $v(x,u)$ (see \p{D0})
as the independent off-shell harmonic field in this action.  However,
their approach produces a perturbatively trivial quantum theory \cite{MOY}.

Let us then deform the bilinear action \p{D5} by terms depending on $\Vp$,
defining a new class of harmonic-field models.
For example, one can choose such terms by analogy with
the harmonic action of $N{=}2$ Yang-Mills theory\cite{Z1}
\be
 S(\Vp)=c\sum\limits^{\infty}_{n=1} {\textstyle{\frac{(-1)^n}{n}}}
\int\!\!d^{\4}\!x\,du_{\s1}\ldots du_n\;
\frac{\Vp(x,u_1)\ldots\Vp(x,u_n)}{(u^+_{\s1}u^+_{\s2})\ldots(u^+_n u^+_{\s1})}
\label{D6}
\ee
where $c$ is a constant of $d{=}{-}4$ and
$(u^+_{\s1} u^+_{\s2} )^{-1}$ is the harmonic distribution \cite{GIOS1}.
The action $S_P + S(\Vp)$ produces the equations
\be
\partial^+_\da \Vp=0 \quad,\qquad
\partial^+_\da P^{\da---}=c\Vm(\Vp) \quad,\lb{D7}
\ee
where the perturbative solution $\Vm(\Vp)$ of the harmonic equation \p{C11}
is to be used. The second equation describes some restriction on the analytic
fields, namely
\be
\int\!\!(d^{\2}\!x)^{--}\;\Vm(\Vp)=0 \quad. \lb{D7b}
\ee

Since the importance of analytical group is not evident
for the off-shell theory, we may consider an arbitrary quadratic term
in the full harmonic space with coordinates $z=(x,u)$:
\be
S_2(\Vp)={\textstyle{{1\over2}}}\int\!\!dz_{\s1}dz_{\s2}\;
K^{--,--}(z_{\s1},z_{\s2})\;\T\;\Vp(z_{\s1})\Vp(z_{\s2}) \quad,\lb{D8}
\ee
where
$K^{--,--}(z_{\s1},z_{\s2})$ is some local or nonlocal kernel (e.g.
$(u^+_{\s1} u^+_{\s2} )^{-2}\delta(x_{\s1}-x_{\s2})$ for eq.\p{D6}).
Higher terms contain polynomials in $\Vp$.
The auxiliary field possesses an additional gauge symmetry,
$\delta P^{\da---}=\pdap a^{----}(x,u)$.
In the quantum version of this model, one can thus introduce the gauge
\be
\partial^-_\da P^{\da---}=0 \quad.\lb{D9}
\ee
Propagators of the harmonic fields in this gauge are
\bea
&& < P^{\da---}(z_{\s1})|\Vp(z_{\s1})>=-{1\over\Box}\pdam_{\s1}
\delta(x_{\s1}{-}x_{\s2})\delta^{--,++)}(u_{\s1},u_{\s2}) \quad,\lb{D10}\\
&& < P^{\da---}(z_{\s1})| P^{\db---}(z_{\s2})>=-{1\over\Box^2}
\pdam_{\s1}\pdbm_{\s2}K^{--,--}(z_{\s1},z_{\s2}) \quad.\lb{D11}
\eea

An action with Lagrange multiplier
can be written also for the nonlinear analyticity equation \p{C14},
\be
S(\Vm,P^{\da+++})=\int\!\!d^{\4}\!x\,du\;\T\;P^{\da+++}
{\cal F}_\da^{---}(\Vm) \quad.\lb{D12}
\ee
Of course, this action can be deformed as well by terms
depending on $\Vm$.

Finally, we discuss shortly a completely different off-shell description of
self-dual harmonic fields, using functional integrals over the
analytic fields $\Vp(\zeta)$ where $\zeta=(x^+_\da,u^\pm_\beta)$.
Candidate actions for this approach can be written directly in the
analytic space. The quadratic term, for instance, have the following form:
\be
\int\!\!d\zeta^{++}_{\s1}d\zeta^{++}_{\s2}\;A^{----,----}
(\zeta_{\s1},\zeta_{\s2})\;\Vp(\zeta_{\s1})\;\Vp(\zeta_{\s2}) \quad,\lb{D13}
\ee
where the analytic kernel relates to
the kernel in the general harmonic space \p{D8} via
\be
A^{----,----}(x^+_{\s1},u^\pm_{\s1},x^+_{\s2},u^\pm_{\s2})=
\int\!\!(d^{\2}\!x_{\s1})^{--} (d^{\2}\!x_{\s2})^{--}\;
K^{--,--}(x^\pm_{\s1},u^\pm_{\s1},x^\pm_{\s2},u^\pm_{\s2}) \quad.
\lb{D14}
\ee
Let us mention that an arbitrary analytic prepotential $\Vp$
describes solutions of the SDYM equations with unbounded value of the
classical action $\int\!\!d^{\4}\!x\,\T (F_{\da\db})^2$. Hence, the relation of
the functional integrals over analytic fields $\Vp(\zeta)$ with the instanton
ideology is unclear.

\setcounter{equation}{0}
\section{\lb{E}  Actions for generalized self-dual equations}
\hspace{0.5cm}

Now we shall consider invariant actions for the generalized self-dual
harmonic equations of sect. \ref{F}.
An action for the first-order equation \p{F2} can be written
by analogy with the non-invariant action \p{B14},
\be
S(P^{+++},\Vm,\Omega^{----})=\int\!\! d^{\4}\!x du\, \T\, P^{\da+++}
\Bigl\{2\partial_\da^-\Vm + [\Vm ,\partial_\da^+\Vm] - \partial_\da^+
\Omega^{----}\Bigr\} + S^\prime(\Vm) \lb{E0}
\ee
where the last term may contain, for instance, polynomial functions of
$\Dp\Vm$.
This action exhibits no invariance with non-analytic parameters and
generates simple Feynman rules.

It is easy to show that the second-order equation \p{F1} can indeed
be derived from a harmonic action with non-covariant measure:
\be
\int\!\!d^{\4}\!x\,du\;f^{++++}(u)\;\T
\Bigl(\Vm\partial^\dap \partial_\da^-\Vm + {\textstyle{{1\over3}}}
\Vm[\partial^{\da+}\Vm,\partial_\da^+\Vm]\Bigr) \quad,\lb{E1}
\ee
where $f^{++++}(u)$ denotes an arbitrary harmonic function, which breaks the
manifest $SU(2)_L$ invariance. This action generates a nontrivial
quantum perturbation theory completely local in harmonic variables.
For instance, the generating functional of the free Green's functions has the
following form:
\be
W(J)=\mbox{ln}Z(J)={1\over2}\int\!\! d^{\4}\!x\, du\; f^{++++}(u)\;\T
\Bigl[ J^\m(x,u) {1\over\Box} J^\m(x,u) \Bigr] \quad,\lb{E2}
\ee
where $J^\m(x,u)$ is the classical source.

The construction of a bilinear action for the harmonic equation $\Box\Vm=0$
is very simple:
\be
S_2(\Vm)={1\over f^2} \int\!\!d^{\4}\!x\,du\;\T\;
(\Dp)^2\Vm\partial^\dap \partial_\da^-\Vm \quad,\lb{E5}
\ee
where $f$ is a coupling constant of $d{=}1$.
Unfortunately, the cubic term local in harmonics should contain an infinite
number of terms, carrying arbitrary powers of the harmonic derivatives $\Dp$
and $\Dm$. We are only in the position to construct these terms iteratively.
The starting trilinear term in this infinite harmonic construction is
\be
{1\over f^2}\int\!\!d^{\4}\!x\,du\;\T\;
(\Dp)^2\Vm[\partial^\dap\Vm, \partial_\da^+\Vm] \quad.\lb{E7}
\ee
The variation of this term produces not only the unique quadratic term in
eq.\p{F1}, but also some additional terms which are undesirable and should
be cancelled by the infinite sum of variations of higher terms.
Thus, we do not know the invariant action for eq.\p{F1} in closed form.
Of course, one may use the term \p{E7} for describing some self-interaction
of $\Vm$ which is not directly connected with the self-duality equation.

Off-shell harmonic fields in our treatment contain
an infinite series of higher-spin fields.
In this approach, the harmonic equation \p{F1} yields an infinite set of
equations for the component fields $V_k^{\alpha_1\cdots\alpha_{2k}}(x)$ \p{D2},
\be
\partial^\dap {\cal F}_\da^{---}(\Vm)
=\sum\limits^{\infty}_{k=1}(u^{+(k-1)} u^{-(k+1)})_{\alpha_1\cdots\alpha_{2k}}
E^{\alpha_1\cdots\alpha_{2k}}_k(V)=0 \quad. \lb{E3}
\ee
Each equation contains the linear Laplace term and the infinite number
of bilinear terms describing the interaction of higher spin fields.
Using the harmonic identities (\ref{G10}-\ref{G14}) one can calculate,
in principle, any term of these equations. Consider, for example, the
lowest terms of the equation with $k=1$,
\bea
&& E^{\alpha_1\alpha_2}_1=\Box V^{\alpha_1\alpha_2}_1 +{\textstyle{{2\over9}}}
[\partial_\rho^\ds V^{\rho\alpha_1}_1\;,\;\partial_{\beta\ds}
V^{\beta\alpha_2}_1]+\nn\\ &&+{\textstyle{{1\over6}}}
[\partial^{(\alpha_1\ds}V_1^{\alpha_2\alpha_3)}~,~
\partial^\beta_\ds V_{1\beta\alpha_3}] -{\textstyle{{1\over10}}}
[\partial^{(\alpha_1\ds}V_1^{\beta_2\beta_3)}~,~
\partial^{(\alpha_2}_\ds V_{1\beta_2\beta_3)}]+ O(V_2) \quad,\lb{E4}
\eea
where all bilinear terms with $k\ge 2$ are omitted.
It is clear that an invariant component action with an infinite number
of bilinear terms exists for the linearized infinite system of equations.
The analysis of the trilinear terms can be performed,
in principle, step by step for interactions of the fields $V_k$.

We shall now demonstrate that the harmonic field $L$ \p{F9}
is in fact more convenient for the construction of an invariant action.
The action (local in $x$ and $u$) for the harmonic non-polynomial equation
\p{F10} can be written by analogy with eq.\p{B9}:
\bea
&&S(L)={1\over f^2} \int\!\!d^{\4}\!x\,du\int\limits^{1}_{0}\!\!ds\;\T\; L\;
\partial^\dap \Bigl[ e^{sL}\;\partial_\da^- e^{-sL} \Bigr] =\nn\\
&&\int\!\!d^{\4}\!x\,du\;\T \Bigl( L\; \partial^\dap \partial_\da^-\;L
-{\textstyle{{1\over3}}}L\;[~\partial^\dap L~,~ \partial_\da^-L~]
+O(L^4) \Bigr) \quad.\lb{E6}
\eea
This action
contains formally non-renormalizable derivative terms and may serve
as effective action for the infinite set of fields
$L^{\alpha_1\cdots\alpha_{4k+2}}(x)$ with odd spins $2k{+}1$.
The action is nondegenerate and does not require gauge fixing for quantization.

Remarkably, the non-covariant field theory for the analogous
action \p{B9} is one-loop finite \cite{LMNS,Ke}.
It is conceivable that a covariant sigma model for the harmonic field $L(x,u)$
has such interesting quantum properties, as well.

\vspace{1.5cm}
\noindent
{\large\bf Acknowledgements}\\
\noindent
B.Z. is grateful to A.S. Galperin and E.A. Ivanov for interesting discussions.
This work is partially supported  by  grants  RFBR-96-02-17634,
RFBR-DFG-96-02-00180,  INTAS-93-127-ext and INTAS-96-0308, by the
Heisenberg-Landau program and
by a grant of the Uzbek Foundation of Basic Research N 11/97.
\vspace{1.5cm}

\setcounter{equation}{0}
\section{\lb{G} Appendix: Relations for $SU(2)/U(1)$ harmonics}
\hspace{0.5cm}
The harmonics $u^\pm_\alpha$ ($\alpha=1,2$) can be interpreted as projective
coordinates of the sphere $SU(2)/U(1)$ \cite{GIKOS,GIOS1}. The basic
relations for the harmonics and the harmonic derivatives are
\bea
&& u^+_\alpha u^-_\beta -u^+_\beta u^-_\alpha=\varepsilon_{\alpha\beta} \quad,
\lb{G1}\\
&& [\dpp,\dm]=\partial^{\s 0}\quad,\qquad
[\partial^{\s 0},\partial^{\s\pm\pm}]=\pm 2\partial^{\s\pm\pm}\quad,\lb{G2}\\
&& \partial^{\s++}\;u^+_\alpha=0\quad,\qquad
   \partial^{\s++}\;u^-_\alpha=u^+_\alpha \quad,\lb{G3}\\
&& \partial^{\s--}\;u^-_\alpha=0\quad,\qquad
   \partial^{\s--}\;u^+_\alpha=u^-_\alpha\quad,\qquad
   \partial^{\s 0}\;u^\pm_\alpha=\pm u^\pm_\alpha\quad. \lb{G4}
\eea

There are two types of conjugation acting on these harmonics. The first
one is the ordinary conjugation
\be
\overline{u^\pm_\alpha}=\pm u^{\alpha\mp}\quad,\qquad
\overline{u^{\alpha\mp}}=\mp u^\pm_\alpha \lb{G5}
\ee
which changes  $U(1)$ charges. The alternative harmonic conjugation
conserves $U(1)$ charges
\be
\widetilde{u^\pm_\alpha}=u^{\alpha\pm}\quad,\qquad
\widetilde{u^{\alpha\pm}}=-u_\alpha^\pm\quad.\lb{G6}
\ee
It should be stressed that the harmonic gauge theory is covariant only
with respect to this $U(1)$-conserving conjugation.
The  $SU(2)_L\times SU(2)_R$ covariant conjugation also acts on
$u$-independent quantities, e.g.
\be
\overline{\varepsilon_{\alpha\rho}}=\varepsilon^{\rho\alpha}\quad,\qquad
\overline{x^\adb}=x_\adb\quad,\qquad
\overline{\pab}=-\partial^\adb\quad.\lb{G7}
\ee
For the case of $SL(2,R)_L\times SU(2,R)_R$ all coordinates and harmonics
are real.

We shall use the notation $(u^{+p}u^{-q})_{\alpha_1,\cdots\alpha_{(p+q)}}$
for the irreducible symmetric combinations of harmonics with $p$ positive
and $q$ negative charges, for example,
\bea
&& (u^+ u^-)_{\alpha_1\alpha_2}=
   {\textstyle{{1\over2}}}(u^+_{\alpha_1} u^-_{\alpha_2}
   +u^+_{\alpha_2} u^-_{\alpha_1})\quad,\lb{G8}\\
&& (u^+ u^{-2})_{\alpha_1\alpha_2\alpha_3}={\textstyle{{1\over3}}}
   (u^+_{\alpha_1} u^-_{\alpha_2} u^-_{\alpha_3}+u^+_{\alpha_2}u^-_{\alpha_3}
   u^-_{\alpha_1} +u^+_{\alpha_3} u^-_{\alpha_1} u^-_{\alpha_2})\quad.\lb{G9}
\eea
Multiplication rules for irreducible harmonics can be obtained with the
help of basic relation \p{G1}:
\bea
&& u^+_\rho ( u^{-2})_{\alpha_1\alpha_2}=
(u^+ u^{-2})_{\rho\alpha_1\alpha_2}+{\textstyle{{2\over3}}}
\varepsilon_{\rho(\alpha_1}u^-_{\alpha_2)} \quad, \lb{G10}\\
&& u^-_\rho (u^+ u^{-2})_{\alpha_1\alpha_2\alpha_3}=
(u^+ u^{-3})_{\rho\alpha_1\alpha_2\alpha_3}-{\textstyle{{1\over4}}}
\varepsilon_{\rho(\alpha_1}(u^{-2})_{\alpha_2\alpha_3)} \quad, \lb{G11}\\
&& u^+_\rho (u^+ u^{-3})_{\alpha_1\alpha_2\alpha_3\alpha_4}=
(u^+ u^{-3})_{\rho\alpha_1\alpha_2\alpha_3\alpha_4}+{\textstyle{{3\over5}}}
\varepsilon_{\rho(\alpha_1}(u^+ u^{-2})_{\alpha_2\alpha_3\alpha_4)} \quad,
\lb{G12}\\
&& u^+_\rho (u^{+2} u^{-4})_{\alpha_1\cdots\alpha_6}=
(u^{+3} u^{-4})_{\rho\alpha_1\cdots\alpha_6}+{\textstyle{{4\over7}}}
\varepsilon_{\rho(\alpha_1}(u^{+2} u^{-3})_{\alpha_2\cdots\alpha_6)} \quad,
\lb{G13}\\
&&(u^+ u^{-2})_{\alpha_1\alpha_2\alpha_3}(u^+ u^{-2})_{\beta_1\beta_2\beta_3}=
(u^{+2}u^{-4})_{\alpha_1\cdots\beta_3}-{\textstyle{{1\over5}}}
\varepsilon_{(\alpha_1(\beta_1}\varepsilon_{\alpha_2\beta_2}
u^-_{\alpha_3)}u^-_{\beta_3)} \quad, \lb{G14}
\eea
where parentheses denote complete symmetrization of indices. Note that in the
last formula we use separate symmetrization of $\alpha_i$ and $\beta_i$
indices.

The harmonic distributions \cite{GIOS1} satisfy the following relations,
\bea
&& \dpp_{\s1} \frac{1}{(u^+_{\s1}u^+_{\s2})^n}=
{\textstyle{{1\over(n-1)}}}(\dm_{\s1})^{n-1} \delta^{(n,-n)}(u_{\s1},u_{\s2})
\quad, \lb{G15}\\
&& \dm_{\s1} \frac{1}{(u^-_{\s1}u^-_{\s2})^n}=
{\textstyle{{1\over(n-1)}}}(\dpp_{\s1})^{n-1} \delta^{(-n,n)}(u_{\s1},u_{\s2})
\quad, \lb{G16}
\eea
where indices 1 and 2 parametrize the independent sets of harmonics.
One can prove directly a useful integral relation for the harmonic
distributions,
\be
\int\!\!du_{\s2}\;\frac{1}{(u^-_{\s1}u^-_{\s2})^2}\;\frac{1}{(u^+_{\s2}
u^+_{\s3})^2}=\delta^{(2,-2)}(u_{\s1},u_{\s3}) \quad. \lb{G17}
\ee

Finally, there exists a useful representation of harmonics in terms of
a real variable $\varphi$ and a complex spectral variable $\lambda$,
\be
\left(\begin{array}{cc} u^-_1 & u^+_1 \\
			u^-_2 & u^+_2 \end{array}\right)=
\frac{1}{\eta(\lambda)}
\left(\begin{array}{cc} 1 &-\lambda\\
		  \bar{\lambda}& 1 \end{array} \right)
\left(\begin{array}{cc} e^{-i\varphi} &0\\
		  0  & e^{i\varphi} \end{array} \right)
\label{G18}
\ee
where $\eta(\lambda)=\sqrt{1+\lambda\bar{\lambda}}$.
The harmonic derivatives then take the following form:
\be
\partial^{\s++}=e^{ 2i\varphi}
[\eta^2(\lambda)\bar{\partial}_{\bar{\lambda}}-(i/2)\lambda\partial_\varphi]
\quad, \label{G19}
\ee
\be
\partial^{\s--}=\;-\;e^{-2i\varphi}
[\eta^2(\lambda)\partial_{\lambda}+(i/2)\bar{\lambda}\partial_\varphi]
\quad, \label{G20}
\ee
\be
\partial^{\s 0}=-i\partial_\varphi
\quad. \label{G21}
\ee

\vfill\eject

\end{document}